\documentclass[11pt]{article}
\usepackage{graphicx}
\usepackage{amssymb}

\title{Notoph-Graviton-Photon Coupling}
\author{V. V. Dvoeglazov\\UAF, Universidad de Zacatecas \\
Apartado Postal 636, Suc. 3 Cruces, Zacatecas 98068, ZAC., M\'exico\\
e-mail: valeri@fisica.uaz.edu.mx}

\date{\empty}

\begin{document}
\maketitle

\begin{abstract}
In the sixties Ogievetski\u{\i} and Polubarinov proposed
the concept of  {\it notoph}, whose helicity
properties are complementary to those of  {\it photon}.
Later, Kalb and Ramond (and others) developed
this theoretical concept. And, at the present times it is widely accepted. 
We analyze the quantum theory of antisymmetric tensor fields
with taking into account mass dimensions of notoph and photon. 
It appears to be  possible the description of both photon and
notoph degrees of freedom on the basis of
the modified Bargmann-Wigner formalism for the symmetric
second-rank spinor. 

Next, we proceed to derive equations for the symmetric tensor
of the second rank on the basis of the Bargmann-Wigner formalism
in a straightforward way. The symmetric multispinor of the fourth
rank is used. It is constructed out of the Dirac 4-spinors.
Due to serious problems with the interpretation of
the results obtained on using the standard procedure we generalize
it, and we obtain the spin-2 relativistic equations, which are consistent
with the general relativity. The importance of the 4-vector field (and its gauge part) 
is pointed out.

Thus, we present the full theory which contains the photon, the notoph (the Kalb-Ramond field)
and the graviton. The relations of this theory with the higher spin theories
are established. In fact, we deduced the gravitational field equations from 
relativistic quantum mechanics. 
The relations of this theory with scalar-tensor theories 
of gravitation and f(R) are discussed. 
We estimate possible interactions, fermion-notoph,
graviton-notoph, photon-notoph, and  we conclude that they will be probably seen 
in experiments  in the next few years. 

PACS number: 03.65.Pm , 04.50.-h , 11.30.Cp
\end{abstract}

\section{Introduction\ldots}\label{s:intro}

In the series of the papers~\cite{DVO,DVO962,DVO961,DVO970,DVO97},
cf. with Refs.~\cite{Ahl,DVA,DVALF}, we tried to find connection between the theory
of the quantized antisymmetric tensor (AST) field of the second rank (and that
of the corresponding 4-vector field) with the $2(2s+1)$ Weinberg-Tucker-Hammer
formalism~\cite{WEIN0,TH1}. 

Several previously published works~\cite{DEB,Og,Hayh,Kalb,Ohanian,Avdeev}, 
introduced the concept of the notoph (the Kalb-Ramond field) which is constructed on the basis of 
the antisymmetric tensor ``potentials". It represents itself the non-trivial 
spin-0 field. The well-known textbooks~\cite{Lurie,Bogol,Novozh,Itzyk} did {\it not}
discuss  the problems, whether the massless quantized AST field
and the quantized 4-vector field are transverse or longitudinal fields
(in the sense if the helicity $h=\pm 1$ or $h=0$)? can the electromagnetic potential 
be a 4-vector in a quantized theory? how should 
the massless limit be taken? and many other fundamental problems of the physics of bosons. 
In my opinion, the most rigorous works are refs.~\cite{BW,WEIN0,Kim,Wein}, but it is not
easy to extract corresponding answers even from them. 

First of all, we note that 1) ``...In natural units
($c=\hbar=1$) ... a lagrangian density, since the action is
dimensionless,
has dimension of [energy]$^4$"; 2) One can always renormalize the
lagrangian density and ``one can obtain the same equations of motion...
by substituting $L \rightarrow (1/M^N) L$, where $M$ is an arbitrary
energy scale",  cf.~\cite{DVO962}; 3) the right physical dimension of
the field strength tensor $F^{\mu\nu}$ is [energy]$^2$; ``the transformation
$F^{\mu\nu} \rightarrow (1/2M) F^{\mu\nu}$ [which was regarded in
Ref.~\cite{DVO97}] ... requires a more detailed study ... [because] the
transformation above changes its physical dimension: it is not a simple
normalization transformation". Furthermore, in the first papers on the
notoph~\cite{Og,Hayh,Kalb}\footnote{It is also known as the longitudinal
Kalb-Ramond field, but the consideration of Ogievetski\u{\i} and Polubarinov
permits to study the $m\rightarrow 0$ procedure.} the authors
used the normalization of the 4-vector $F^\mu$ field\footnote{It is well
known that it is related to the third-rank antisymmetric field tensor.} to
[energy]$^2$ and, hence, the antisymmetric tensor ``potentials"
$A^{\mu\nu}$, to [energy]$^1$. We try to discuss these problems on the basis of the
generalized Bargmann-Wigner formalism~\cite{BW}. Thus, the Proca and Maxwell formalisms
are generalized, see, e.~g., Ref.~\cite{DVO-HJS}.

In the Sections 3 and 4 we consider the spin-2 equations. 
The general scheme for derivation of higher-spin equations
has been given in~\cite{BW}. A field of the rest mass $m$ and the spin $s \geq
{1\over 2}$ is represented by a completely symmetric multispinor of rank $2s$.
The particular cases $s=1$ and $s={3\over 2}$ have been considered in the
textbooks, e.~g., Ref.~\cite{Lurie}. The spin-2 case can also be of some
interest because we can believe that the essential features of
the gravitational field are  obtained from transverse components of the
$(2,0)\oplus (0,2)$  representation of the Lorentz group. Nevertheless,
questions of the redundant components of the higher-spin relativistic
equations are not yet understood in detail~\cite{Kirch}.

In the last Sections we discuss the questions of interactions.

\section{Photon-Notoph Equations.}

For spin 1 we start from\footnote{In the classic works on this formalism the authors worked 
in the Euclidean metrics. However, there is no any problem to write
the equations and other formulas in the pseudo-Euclidean metrics accustomed today;
just change the sign of $p_\mu p_\mu$, and other products.}
\begin{equation}
[\gamma_{\alpha\beta} p_\alpha p_\beta +A p_\alpha p_\alpha +Bm^2] \Psi
=0\,, \label{wth1}\end{equation} where $p_\mu=-i\partial_\mu$ and
$\gamma_{\alpha\beta}$ are the Barut-Muzinich-Williams 
covariantly - defined $6\times 6$ matrices, $\sum_{\mu}^{}
\gamma_{\mu\mu}=0$.  The determinant of $[\gamma_{\alpha\beta} p_\alpha
p_\beta +A p_\alpha p_\alpha +Bm^2]$ is of the 12th order in $p_\mu$.
If we are interested in solutions with $E^2 -{\bf p}^2 = m^2$, $c=\hbar =1$, 
they can be obtained on using the constraints in the above equation:
\begin{equation} 
\frac{B}{A+1} =1\,,\quad \frac{B}{A-1}=1\,.
\end{equation}
We may also have the tachyonic solutions, etc.
The particular cases are:
\begin{itemize}

\item
$A=0, B=1$ $\Leftrightarrow$ we have the Weinberg's equation for $s=1$
with 3 solutions $E=+\sqrt{{\bf p}^2 +m^2}$, 3 solutions
$E=-\sqrt{{\bf p}^2 +m^2}$, 3 solutions $E=+\sqrt{{\bf p}^2 -m^2}$
and 3 solutions $E=-\sqrt{{\bf p}^2 -m^2}$. Tachyonic solutions have been reformulated in various ways,
for instance, as the ones leading to the spontaneous symmetry breaking,
and to the non-zero quantum  vacuum.

\item
$A=1, B=2$ $\Leftrightarrow$ we have the Tucker-Hammer equation for $s=1$.
The solutions are 
with $E=\pm\sqrt{{\bf p}^2 +m^2}$ {\it only}.

\end{itemize}

Thus, the addition of the Klein-Gordon equation to (\ref{wth1}) may change the physical content even on the {\it free} level.

What are the corresponding equations for the antisymmetric tensor field? They can be the Proca equations in the massive case, and the 
Maxwell equations in the massless case. We have shown in Refs.~\cite{DVO,DVO962}
that one can obtain {\it four} different equations
for antisymmetric tensor fields from the Weinberg's $2(2s+1)-$
component formalism.  First of all, we note that $\Psi$ is, in fact,
bivector, ${\bf E}_i = -iF_{4i}$, ${\bf B}_i = {1\over 2} \epsilon_{ijk}
F_{jk}$,, or ${\bf E}_i = -{1\over 2} \epsilon_{ijk} \tilde F_{jk}$, ${\bf
B}_i = -i \tilde F_{4i}$, or their combinations.  One can separate
the four cases:
\begin{itemize}

\item
$\Psi^{(I)} = \pmatrix{{\bf E} +i{\bf B}\cr
{\bf E} -i{\bf B}\cr}$, $P=-1$, where ${\bf E}_i$ and ${\bf B}_i$ are the
components of the tensor.

\item
$\Psi^{(II)} = \pmatrix{{\bf B} -i{\bf E}\cr
{\bf B} +i{\bf E}\cr}$, $P=+1$, where ${\bf
E}_i$, ${\bf B}_i$ are the components of the tensor.

\item
$\Psi^{(III)} = \Psi^{(I)}$, but (!) ${\bf E}_i$ and ${\bf B}_i$
are the
corresponding vector and axial-vector  components of the
{\it dual} tensor \,\,\, $\tilde F_{\mu\nu}$.

\item
$\Psi^{(IV)} = \Psi^{(II)}$, where ${\bf E}_i$ and ${\bf B}_i$
are the components of the {\it dual} tensor $\tilde F_{\mu\nu}$.

\end{itemize}
The mappings of the WTH equations are:
\begin{eqnarray}
&&\partial_\alpha\partial_\mu F_{\mu\beta}^{(I)}
-\partial_\beta\partial_\mu F_{\mu\alpha}^{(I)}
+ {A-1\over 2} \partial_\mu \partial_\mu F_{\alpha\beta}^{(I)}
-{B\over 2} m^2 F_{\alpha\beta}^{(I)} = 0\,,\label{wth1a}\\
&&\partial_\alpha\partial_\mu F_{\mu\beta}^{(II)}
-\partial_\beta\partial_\mu F_{\mu\alpha}^{(II)}
- {A+1\over 2} \partial_\mu \partial_\mu F_{\alpha\beta}^{(II)}
+{B\over 2} m^2 F_{\alpha\beta}^{(II)} = 0\,,\\
&&\partial_\alpha\partial_\mu \tilde F_{\mu\beta}^{(III)}
-\partial_\beta\partial_\mu \tilde F_{\mu\alpha}^{(III)}
- {A+1\over 2} \partial_\mu \partial_\mu \tilde F_{\alpha\beta}^{(III)}
+{B\over 2} m^2 \tilde F_{\alpha\beta}^{(III)} = 0\,,\nonumber\\
&&\\
&&\partial_\alpha\partial_\mu \tilde F_{\mu\beta}^{(IV)}
-\partial_\beta\partial_\mu \tilde F_{\mu\alpha}^{(IV)}
+ {A-1\over 2} \partial_\mu \partial_\mu \tilde F_{\alpha\beta}^{(IV)}
-{B\over 2} m^2 \tilde F_{\alpha\beta}^{(IV)} = 0\,.
\end{eqnarray}
In the Tucker-Hammer case ($A=1, B=2$) we can recover the Proca theory
from (\ref{wth1a}):
\begin{equation}
\partial_\alpha \partial_\mu F_{\mu\beta}
-\partial_\beta \partial_\mu F_{\mu\alpha} = m^2 F_{\alpha\beta}
\label{proca1}\,,
\end{equation}
($A_\nu ={1\over m^2}  \partial_\alpha F_{\alpha\nu}$ should be substituted in
$F_{\mu\nu} = \partial_\mu A_\nu -\partial_\nu A_\mu$, and the result is multiplied by
$m^2$).

We also noted that the massless limit of this theory does {\it not} coincide
with  the Maxwell theory in some cases, while it contains the latter as a particular
case.  In~\cite{DVO961,DVO97,Dvo97} we showed that it is possible to define
various massless limits for the Duffin-Kemmer-Proca theory. Another one is
the Ogievetski\u{\i}-Polubarinov {\it notoph} (which is also called 
the Kalb-Ramond field), Ref.~\cite{Og} in the US 
literature. The transverse
components of the AST field can be removed from the corresponding
Lagrangian by means of the ``new gauge transformation" $A_{\mu\nu}
\rightarrow A_{\mu\nu} +\partial_\mu \Lambda_\nu -\partial_\nu
\Lambda_\mu$, with the vector gauge function $\Lambda_\mu$.

The second (II) case is
\begin{equation}
\partial_\alpha\partial_\mu F_{\mu\beta} -\partial_\beta \partial_\mu F_{\mu\alpha}
= [\partial_\mu \partial_\mu - m^2] F_{\alpha\beta}\,.
\end{equation}
So, on the mass shell we have $[\partial_\mu \partial_\mu - m^2] F_{\alpha\beta} =0$,
and, hence,
\begin{equation}
\partial_\alpha\partial_\mu F_{\mu\beta} -\partial_\beta \partial_\mu
F_{\mu\alpha} = 0\,,\label{str}
\end{equation}
which rather corresponds to the Maxwell-like case. However, if
we calculate dispersion relations for the second case, Eq. (\ref{str}), it
appears that the equation has solutions even if $m\neq 0$.

Now we are interested in the {\it parity-violating}
equations for antisymmetric tensor fields. We  investigate
the most general mapping of the Weinberg-Tucker-Hammer formulation
to the antisymetric tensor field formulation too.
Instead of $\Psi^{(I-IV)}$ we shall try to use now
\begin{equation}
\Psi^{(A)} = \pmatrix{{\bf E} +i{\bf B}\cr
{\bf B} +i{\bf E}\cr} = {1+\gamma^5\over 2} \Psi^{(I)}+
{1-\gamma^5 \over 2} \Psi^{(II)}\,.
\end{equation}
As a result, the equation for the AST fields is
\begin{equation}
\partial_\alpha \partial_\mu F_{\mu\beta}
-\partial_\beta \partial_\mu F_{\mu\alpha}
={1\over 2} (\partial_\mu \partial_\mu) F_{\alpha\beta} +
[-{A\over 2} (\partial_\mu \partial_\mu) + {B\over 2} m^2] \tilde
F_{\alpha\beta}\,.\label{pv1}
\end{equation}
Of course, $\Psi^{(A)^\prime}
=\pmatrix{{\bf B} - i{\bf E}\cr {\bf E} -i{\bf B}\cr} = -i \Psi^{(A)}$,
and the equation is unchanged.
The different choice is
\begin{equation} \Psi^{(B)} = \pmatrix{{\bf E} +i{\bf
B}\cr -{\bf B} -i{\bf E}\cr} = {1+\gamma^5\over 2} \Psi^{(I)}- {1-\gamma^5
\over 2} \Psi^{(II)}\,.
\end{equation}
Thus, one has
\begin{equation}
\partial_\alpha \partial_\mu F_{\mu\beta}
-\partial_\beta \partial_\mu F_{\mu\alpha}
={1\over 2} (\partial_\mu \partial_\mu) F_{\alpha\beta} +
[{A\over 2} (\partial_\mu \partial_\mu)- {B\over 2} m^2] \tilde
F_{\alpha\beta}\,.\label{pv2}
\end{equation}
Of course, one can also
use the dual tensor (${\bf E}^i = -{1\over 2} \epsilon_{ijk} \tilde
F_{jk}$ and ${\bf B}^i =-i\tilde F_{4i}$) and obtain analogous equations:
\begin{eqnarray}
&&\partial_\alpha \partial_\mu \tilde F_{\mu\beta}
-\partial_\beta \partial_\mu \tilde F_{\mu\alpha}
={1\over 2} (\partial_\mu \partial_\mu) \tilde F_{\alpha\beta} +
[-{A\over 2} (\partial_\mu \partial_\mu) + {B\over 2} m^2]
F_{\alpha\beta}\,,\nonumber\\
&&\\
&&\partial_\alpha \partial_\mu \tilde F_{\mu\beta}
-\partial_\beta \partial_\mu \tilde F_{\mu\alpha}
={1\over 2} (\partial_\mu \partial_\mu) \tilde F_{\alpha\beta} +
[{A\over 2} (\partial_\mu \partial_\mu) - {B\over 2} m^2]
F_{\alpha\beta}\,. \nonumber
\\ 
\end{eqnarray}
They are connected
with (\ref{pv1},\ref{pv2}) by the dual transformations.

The states corresponding to the new functions $\Psi^{(A)}$,
$\Psi^{(B)}$ etc are {\it not} the parity eigenstates. So, it is not
surprising that we have $F_{\alpha\beta}$ and
its dual $\tilde F_{\alpha\beta}$ in
the same equations. In total we have already eight equations.

One can also consider the most general case
\begin{equation}
\Psi^{(W)} =\pmatrix{aF_{4i} +b \tilde F_{4i} +c \epsilon_{ijk} F_{jk}
+d \epsilon_{ijk} \tilde F_{jk}\cr
eF_{4i} +f \tilde F_{4i} +g \epsilon_{ijk} F_{jk}
+h \epsilon_{ijk} \tilde F_{jk}\cr}\,.
\end{equation}
So, we shall have dynamical equations for $F_{\alpha\beta}$
and $\tilde F_{\alpha\beta}$ with additional parameters $a,b,c,d,\ldots
\in\, {\bf C}$. We have a lot of antisymmetric tensor fields here.
However,
\begin{itemize}

\item
the covariant form preserves if there are some restrictions on the
parameters, only. Alternatively, we have some additional terms of $\partial_4^2$
or ${\bf \nabla}^2$;

\item
both $F_{\mu\nu}$ and $\tilde F_{\mu\nu}$ are present in the equations;

\item
under the definite choice of $a,b,c,d\ldots$ the
equations can be reduced
to the above equations
for the tensor ${\cal H}_{\mu\nu}$ and its dual:
\begin{equation}
{\cal H}_{\mu\nu} = c_1 F_{\mu\nu} +c_2 \tilde F_{\mu\nu} +
{ c_3\over 2} \epsilon_{\mu\nu\alpha\beta} F_{\alpha\beta}
+{c_4 \over 2} \epsilon_{\mu\nu\alpha\beta} \tilde F_{\alpha\beta}\,;
\end{equation}

\item
the parity properties of $\Psi^{(W)}$ are very complicated.
\end{itemize}

Anther way of constructing the  equations of high-spin particles has been given
in~\cite{BW,Lurie}.\footnote{On can also obtain the $s=0$ Kemmer equations on using
the Bargmann-Wigner procedure. One should use the {\it antisymmetric} second-rank multispinor
in this case.}
Bargmann and Wigner claimed
explicitly that they constructed $(2s+1)$ states.\footnote{The
Weinberg-Tucker-Hammer theory has  essentially $2(2s+1)$
components.}  Below we present the
standard Bargmann-Wigner formalism for the spin $s=1$ (and turn to the standard 
pseudo-Euclidean metric):
\begin{eqnarray} \left [ i\gamma^\mu \partial_\mu - m \right
]_{\alpha\beta} \Psi_{\beta\gamma} &=& 0\,,\label{bw1}\\ \left [
i\gamma^\mu \partial_\mu - m \right ]_{\gamma\beta} \Psi_{\alpha\beta} &=&
0\,, \label{bw2} \end{eqnarray}
If one has
\begin{equation} \Psi_{\left \{ \alpha\beta
\right \} } = (\gamma^\mu R)_{\alpha\beta} A_\mu +
(\sigma^{\mu\nu} R)_{\alpha\beta} F_{\mu\nu}\,,
\end{equation} 
with\footnote{The reflection operator $R$ has the
properties
\begin{eqnarray}
&& R^T = -R\,,\quad R^\dagger =R =
R^{-1}\,,\\ && R^{-1} \gamma^5 R = (\gamma^5)^T\,,\\ && R^{-1}
\gamma^\mu R = -(\gamma^\mu)^T\,,\\ && R^{-1} \sigma^{\mu\nu} R = -
(\sigma^{\mu\nu})^T\,.
\end{eqnarray}}
\begin{equation} R = e^{i\varphi}
\pmatrix{\Theta&0\cr 0&-\Theta\cr}\,\quad
\Theta=\pmatrix{0&-1\cr
1&0\cr}
\end{equation} in the spinorial
representation of $\gamma$-matrices, we obtain
the Duffin-Kemmer-Proca equations:
\begin{eqnarray}
&&\partial^\alpha F_{\alpha\mu} = {m\over 2} A_\mu\,,\\
&& 2m F_{\mu\nu} = \partial_\mu A_\nu - \partial_\nu A_\mu\,.
\end{eqnarray}
In order to obtain these equations one should add the equations
(\ref{bw1},\ref{bw2}) and compare functional coefficients at the
corresponding commutators, see Ref.~\cite{Lurie}.  After the corresponding
re-normalization $A_\mu \rightarrow 2m A_\mu$ (or $F^{\mu\nu}\rightarrow (1/2m) F^{\mu\nu}$), 
we obtain the standard textbook set:
\begin{eqnarray} &&\partial^\alpha
F_{\alpha\mu} = m^2 A_\mu\,,\\ && F_{\mu\nu} = \partial_\mu A_\nu -
\partial_\nu A_\mu\,.  \end{eqnarray}
It gives the equation (\ref{proca1})
for the antisymmetric tensor field. Of course, one can investigate 
other sets of equations with different normalization of the $F_{\mu\nu}$ and $A_\mu$ fields. 
Are all these sets of equations equivalent? As we see, to answer this question is not
trivial. It was argued that the physical normalization is such that in the
massless limit the zero-momentum field functions should vanish in the momentum
representation (there are no massless particles at rest). Moreover, we advocate 
the following approach: the massless limit can and must be taken in the end of all
calculations only, i. e., for physical quantities.

How can one obtain other equations
following from the Weinberg-Tucker-Hammer approach?
The recipe for the third equation is simple: use, instead of
$(\sigma^{\mu\nu} R) F_{\mu\nu}$, another symmetric matrix $(\gamma^5
\sigma^{\mu\nu} R)  F_{\mu\nu}$.

After taking into account the above observations let us repeat the procedure
of derivation of the Proca equations from the Bargmann-Wigner
equations for a {\it symmetric} second-rank spinor. However, we now use
\begin{equation}
\Psi_{\{\alpha\beta\}} = (\gamma^\mu R)_{\alpha\beta} (c_a m A_\mu +
c_f  F_\mu) +(\sigma^{\mu\nu} R)_{\alpha\rho} (c_A m (\gamma^5)_{\rho\beta}
A_{\mu\nu} + c_F I_{\rho\beta} F_{\mu\nu})\, ,\label{si}
\end{equation}
with the same $R$ and $\Theta$ as above.
%\begin{equation}
%R=\pmatrix{i\Theta & 0\cr 0&-i\Theta\cr} \quad,\quad \Theta = -i\sigma_2
%= \pmatrix{0&-1\cr 1&0\cr}\, .
%\end{equation}
Matrices $\gamma^{\mu}$ are again chosen in the Weyl (spinorial) representation, i.e.,
$\gamma^5$ is assumed to be diagonal.  Constants $c_i$ are some
numerical dimensionless coefficients. 
The properties of the reflection operator $R$ are  necessary for the expansion (\ref{si}) to be possible in such
a form, i.e., in order to have the $\gamma^{\mu} R$, $\sigma^{\mu\nu} R$
and $\gamma^5 \sigma^{\mu\nu} R$ to be {\it symmetric} matrices.

The substitution of the above expansion into the Bargmann-Wigner
equations, Ref.~\cite{Lurie},
gives us the new Proca-like equations:
\begin{eqnarray}
&&c_a m (\partial_\mu A_\nu - \partial_\nu A_\mu ) +
c_f (\partial_\mu F_\nu -\partial_\nu F_\mu ) =
ic_A m^2 \epsilon_{\alpha\beta\mu\nu} A^{\alpha\beta} +
2 m c_F F_{\mu\nu} \, , \label{pr1} \\
&&c_a m^2 A_\mu + c_f m F_\mu =
i c_A m \epsilon_{\mu\nu\alpha\beta} \partial^\nu A^{\alpha\beta} +
2 c_F \partial^\nu F_{\mu\nu}\, . \label{pr2}
\end{eqnarray}
In the case $c_a=1$, $c_F ={1\over 2}$ and $c_f=c_A=0$ they
are reduced to the ordinary Proca equations.\footnote{We still note
that the division by $m$ in the first equation
is {\it not} the well-defined operation in the case if someone
is interested in the subsequent limiting procedure $m\rightarrow 0$.
Probably, in order to avoid this obscure point one may wish
to write the Dirac equations in the form $\left [ (i\gamma^\mu
\partial_\mu)/m - I \right ] \psi (x) =0$, which
follows straightforwardly in the derivation of the Dirac equation
on the basis of the Ryder relation~\cite{DVA}  and the Wigner
rules for the boosts of the field functions from the zero-momentum frame.}
In the general case we obtain
dynamical equations which connect the photon, the notoph and
their potentials. The divergent (in $m\rightarrow 0$) parts
of field functions and those of dynamical variables
should be removed by corresponding gauge (or Kalb-Ramond gauge)
transformations. It is well known that the notoph massless field is
considered to be the pure
longitudinal field after one takes into account $\partial_\mu
A^{\mu\nu}=
0$.  Apart from these dynamical equations we can obtain a number of
constraints by means of the subtraction of the equations of the
Bargmann-Wigner system (instead of the addition as for
(\ref{pr1},\ref{pr2})). They read
\begin{eqnarray}
&&mc_a \partial^\mu A_\mu + c_f \partial^\mu f_\mu =0\, , \\
&&mc_A \partial^\alpha A_{\alpha\mu} + {i\over 2}
c_F \epsilon_{\alpha\beta\nu\mu}
\partial^\alpha F^{\beta\nu} = 0,\, 
\end{eqnarray}
that suggests $\widetilde F^{\mu\nu} \sim im A^{\mu\nu}$
and $f^\mu \sim mA^\mu$, as in~\cite{Og}.

Thus, after the suitable choice of the dimensionless coefficients
$c_i$ the
Lagrangian density for the photon-notoph field can be proposed:
\begin{eqnarray}
{\cal L} &=& {\cal L}^{Proca} +{\cal L}^{Notoph} = -
{1\over 8} F_\mu F^\mu -{1\over 4} F_{\mu\nu} F^{\mu\nu} +\nonumber\\
&+& {m^2 \over 2} A_\mu A^\mu + {m^2 \over 4} A_{\mu\nu} A^{\mu\nu}\, ,
\end{eqnarray}
The limit $m\rightarrow 0$ may be taken for dynamical variables,
in the end of calculations only.

Furthermore, it is logical to introduce the normalization scalar field
$\varphi (x)$, and consider the expansion:
\begin{equation}
\Psi_{\{\alpha\beta\}} = (\gamma^\mu R)_{\alpha\beta} (\varphi A_\mu)
+ (\sigma^{\mu\nu} R)_{\alpha\beta} F_{\mu\nu}\, .
\end{equation}
Then, we arrive at the following set
\begin{eqnarray}
&&2m F_{\mu\nu} = \varphi (\partial_\mu A_\nu -\partial_\nu A_\mu)
+ (\partial_\mu \varphi) A_\nu - (\partial_\nu \varphi) A_\mu\, ,\\
&& \partial^\nu F_{\mu\nu} = {m \over 2} (\varphi A_\mu)\, ,
\end{eqnarray}
which in the case of the constant
scalar field $\varphi = 2m$  can also be reduced to the system of the
Proca equations. The additional constraints are
\begin{eqnarray}
&&(\partial^\mu \varphi) A_\mu + \varphi (\partial^\mu A_\mu) =0\,,\\
&& \partial_\mu \widetilde F^{\mu\nu} = 0\, .
\end{eqnarray}
 
At the moment it is not yet obvious, how can we
account for other equations in the $(1,0)\oplus (0,1)$
representation, e.g.~[7b], rigorously.   For instance, one can wish to seek the generalization of
the Proca equations on the basis of the introduction of two mass parameters
$m_1$ and $m_2$. But, when we apply the BW procedure to the Dirac
equations we cannot obtain new physical content.  Another equation in the
$(1/2,0)\oplus (0,1/2)$ representation was discussed in
Ref.~\cite{Raspini}.  It has the form:
\begin{equation} \left [
i\gamma^\mu \partial_\mu - m_1 - \gamma^5 m_2 \right ] \Psi (x) =0\,.
\end{equation}
The Bargmann-Wigner procedure for this system of 
equations (which include the $\gamma^5$ matrix in the mass term) yields:
\begin{eqnarray}.
&&2m_1  F^{\mu\nu} +2i m_2 \widetilde F^{\mu\nu} = \varphi
(\partial^\mu A^\nu -\partial^\nu A^\mu) +
(\partial^\mu \varphi) A^\nu - (\partial^\nu \varphi) A^\mu\, ,\\
&&\partial^\nu F_{\mu\nu} = {m_1 \over 2} (\varphi A_\mu), \,
\end{eqnarray}
with the constraints
\begin{eqnarray}
&&(\partial^\mu \varphi) A_\mu +\varphi (\partial^\mu A_\mu) = 0\,, \\
&&\partial^\nu \widetilde F_{\mu\nu} = {im_2 \over 2} (\varphi A_\mu)\,.
\end{eqnarray}
In general, we can now use the four different mass parameters in the equations which are
analogous to (\ref{bw1},\ref{bw2}). However,
the equality of mass factors\footnote{Here, the superscripts $(1)$ and $(2)$ refers to the first and the second equations, respectively, in the modified Bargmann-Wigner system.} ($m_1^{(1)} = m_1^{(2)}$
and $m_2^{(1)} = m_2^{(2)}$) 
is obtained  as necessary conditions in the process of calculations in the system of the
Dirac-like equations. 

In fact, the results of this paper develop the
old results of Ref.~\cite{Og}.   According
to~\cite[Eqs.(9,10)]{Og}
we proceed in  constructing the ``potentials" for the notoph as
follows:\footnote{The notation is that of Ref.~\cite{Og} here.}
\begin{equation} A_{\mu\nu} ({\bf p})  = N \left
[\epsilon_\mu^{(1)} ({\bf p})\epsilon_\nu^{(2)} ({\bf p})-
\epsilon_\nu^{(1)} ({\bf p}) \epsilon_\mu^{(2)} ({\bf p}) \right ]\,.\label{definition}
\end{equation} 
We use explicit forms for the polarization vectors  (e.g., Refs.~\cite{Wein}
and~\cite[formulas(15a,b)]{DVO97}) boosted to the momentum ${\bf p}$:
\begin{eqnarray}
\epsilon^\mu  ({\bf 0}, +1) = - {1\over \sqrt{2}}
\pmatrix{0\cr 1\cr i \cr 0\cr}\,,
\epsilon^\mu  ({\bf 0}, 0) =
\pmatrix{0\cr 0\cr 0 \cr 1\cr}\,,
\epsilon^\mu  ({\bf 0}, -1) = {1\over \sqrt{2}}
\pmatrix{0\cr 1\cr -i \cr 0\cr}\,,
\end{eqnarray}
and ($\widehat p_i = p_i /\vert {\bf p} \vert$,\, $\gamma
= E_p/m$), Ref.~\cite[p.68]{Wein} or Ref.~\cite[p.108]{Novozh},
\begin{eqnarray} 
&& \epsilon^\mu ({\bf p}, \sigma) =
L^{\mu}_{\quad\nu} ({\bf p}) \epsilon^\nu ({\bf 0},\sigma)\,,\\ 
&& L^0_{\quad 0} ({\bf p}) = \gamma\, ,\quad L^i_{\quad 0} ({\bf p}) =
L^0_{\quad i} ({\bf p}) = \widehat p_i \sqrt{\gamma^2 -1}\,,\\
&& L^i_{\quad k} ({\bf p}) = \delta_{ik} +(\gamma -1) \widehat p_i \widehat
p_k\,.
\end{eqnarray}
$N$, the normalization factor, should be taken into account for possible
analyses of propagators and massless limits. After 
substitutions in the definition (\ref{definition})
one obtains
\begin{eqnarray} A^{\mu\nu} ({\bf p}) = {iN^2 \over m} \pmatrix{0&-p_2&
p_1& 0\cr p_2 &0& m+{p_r p_l\over p_0+m} & {p_2 p_3\over p_0 +m}\cr -p_1
&-m - {p_r p_l \over p_0+m}& 0& -{p_1 p_3\over p_0 +m}\cr 0& -{p_2 p_3
\over p_0 +m} & {p_1 p_3 \over p_0+m}&0\cr}\, , \label{lc}
\end{eqnarray}
i.e., it coincides with the longitudinal components of the antisymmetric
tensor obtained in Refs.~[7a,Eqs.(2.14,2.17)]
and~\cite[Eqs.(17b,18b)]{DVO97}  within the normalization and
different forms of the spin basis.  The $A_{\mu\nu} ({\bf p})$
potential reduces to zero in the
limiting case ($m\rightarrow 0$) under appropriate choice of the normalization 
$N=m^\alpha$, $\alpha > 1/2$. If $N=\sqrt{m}$ this reduction of the non-transverse state
occurs if a $s=1$ particle moves along with the third axis $OZ$.\footnote{But, even in this case we cannot have
a good behaviour of the 4-vector fields/potentials in the massless limit in the instant form of the relativistic dynamics,
cf.~\cite{DVALF}.} It is also
useful to compare Eq. (\ref{lc}) with the formula (B2) in
Ref.~\cite{DVALF}
in order to think about correct procedures for taking the massless limits.

Next, the Tam-Happer experiments~\cite{TH} on two laser beams interaction did not
find satisfactory explanation in the framework of the ordinary QED (at
least, their explanation is complicated by huge technical calculations).
On the other hand, in Ref.~\cite{Pradhan} a very interesting model has
been proposed.  It is based on
gauging the Dirac field on using the coordinate-dependent parameters
$\alpha_{\mu\nu} (x)$ in
\begin{equation}
\psi(x) \rightarrow \psi^\prime (x^\prime) = \Omega \psi(x)\,\,, \quad
\Omega = \exp \left [ {i\over 2} \sigma^{\mu\nu} \alpha_{\mu\nu}(x)
\right ]\, ,
\end{equation}
and, thus,  the second ``photon" was introduced. The
compensating 24-component (in general) field $B_{\mu,\nu\lambda}$
reduces
to the 4-vector field as follows (the notation of~\cite{Pradhan} is used
here):
\begin{equation}
B_{\mu,\nu\lambda} = {1\over 4} \epsilon_{\mu\nu\lambda\sigma} a_\sigma
(x) \, .
\end{equation}
As readily seen, after comparison of these formulas with those of
Refs.~\cite{Og,Hayh,Kalb}, the second photon is nothing more than the
Ogievetski\u{\i}-Polubarinov {\it notoph} within the normalization.
Parity properties are dependent not only on the explicit forms of the mo\-men\-tum-space 
field functions of the $(1/2,1/2)$ representation,
but also on the properties of corresponding creation/annihilation
operators. Helicity properties depend on the normalization.

\section{The Standard Bargmann-Wigner Formalism Applied for Spin 2.}

In this Section we use the commonly-accepted procedure
for the derivation  of higher-spin equations~\cite{BW}. 
We begin with the equations for the 4-rank symmetric spinor:
\begin{eqnarray}
\left [ i\gamma^\mu \partial_\mu - m \right ]_{\alpha\alpha^\prime}
\Psi_{\alpha^\prime \beta\gamma\delta} &=& 0\, ,\\
\left [ i\gamma^\mu \partial_\mu - m \right ]_{\beta\beta^\prime}
\Psi_{\alpha\beta^\prime \gamma\delta} &=& 0\, ,\\
\left [ i\gamma^\mu \partial_\mu - m \right ]_{\gamma\gamma^\prime}
\Psi_{\alpha\beta\gamma^\prime \delta} &=& 0\, ,\\
\left [ i\gamma^\mu \partial_\mu - m \right ]_{\delta\delta^\prime}
\Psi_{\alpha\beta\gamma\delta^\prime} &=& 0\, .
\end{eqnarray}  
The massless limit (if one needs) should be taken in the end of all
calculations.

We proceed expanding the field function in the set of symmetric matrices
(as in the spin-1 case, cf.~Ref.~\cite{DVO97}). In the beginning let us use the
first two indices:\footnote{The matrix $R$ can be related to the
$CP$ operation in the $(1/2,0)\oplus (0,1/2)$ representation.}
\begin{equation} \Psi_{\{\alpha\beta\}\gamma\delta} =
(\gamma_\mu R)_{\alpha\beta} \Psi^\mu_{\gamma\delta}
+(\sigma_{\mu\nu} R)_{\alpha\beta} \Psi^{\mu\nu}_{\gamma\delta}\, .
\label{bwff}
\end{equation}
We would like to write
the corresponding equations for functions $\Psi^\mu_{\gamma\delta}$
and $\Psi^{\mu\nu}_{\gamma\delta}$ in the form:
\begin{eqnarray}
&&{2\over m} \partial_\mu \Psi^{\mu\nu}_{\gamma\delta} = -
\Psi^\nu_{\gamma\delta}\, , \label{p1}\\
&&\Psi^{\mu\nu}_{\gamma\delta} = {1\over 2m}
\left [ \partial^\mu \Psi^\nu_{\gamma\delta} - \partial^\nu
\Psi^\mu_{\gamma\delta} \right ]\, \label{p2}.
\end{eqnarray}  
Constraints $(1/m) \partial_\mu \Psi^\mu_{\gamma\delta} =0$
and $(1/m) \epsilon^{\mu\nu}_{\quad\alpha\beta}\, \partial_\mu
\Psi^{\alpha\beta}_{\gamma\delta} = 0$ can be regarded as the consequence
of
Eqs.  (\ref{p1},\ref{p2}).

Next, we present the vector-spinor and tensor-spinor functions as
\begin{eqnarray}
&&\Psi^\mu_{\{\gamma\delta\}} = (\gamma^\kappa R)_{\gamma\delta}
G_{\kappa}^{\quad \mu} +(\sigma^{\kappa\tau} R )_{\gamma\delta}
F_{\kappa\tau}^{\quad \mu} \, ,\\
&&\Psi^{\mu\nu}_{\{\gamma\delta\}} = (\gamma^\kappa R)_{\gamma\delta}
T_{\kappa}^{\quad \mu\nu} +(\sigma^{\kappa\tau} R )_{\gamma\delta}
R_{\kappa\tau}^{\quad \mu\nu} \, ,
\end{eqnarray} 
i.~e.,  using the symmetric matrix coefficients in indices $\gamma$ and
$\delta$. Hence, the total function is
\begin{eqnarray}
\lefteqn{\Psi_{\{\alpha\beta\}\{\gamma\delta\}}
= (\gamma_\mu R)_{\alpha\beta} (\gamma^\kappa R)_{\gamma\delta}
G_\kappa^{\quad \mu} + (\gamma_\mu R)_{\alpha\beta} (\sigma^{\kappa\tau}
R)_{\gamma\delta} F_{\kappa\tau}^{\quad \mu} + } \nonumber\\
&+& (\sigma_{\mu\nu} R)_{\alpha\beta} (\gamma^\kappa R)_{\gamma\delta}
T_\kappa^{\quad \mu\nu} + (\sigma_{\mu\nu} R)_{\alpha\beta}
(\sigma^{\kappa\tau} R)_{\gamma\delta} R_{\kappa\tau}^{\quad\mu\nu} \,,
\end{eqnarray}
and the resulting tensor equations are:
\begin{eqnarray}
&&{2\over m} \partial_\mu T_\kappa^{\quad \mu\nu} =
-G_{\kappa}^{\quad\nu}\, ,\\
&&{2\over m} \partial_\mu R_{\kappa\tau}^{\quad \mu\nu} =
-F_{\kappa\tau}^{\quad\nu}\, ,\\
&& T_{\kappa}^{\quad \mu\nu} = {1\over 2m} \left [
\partial^\mu G_{\kappa}^{\quad\nu}
- \partial^\nu G_{\kappa}^{\quad \mu} \right ] \, ,\\
&& R_{\kappa\tau}^{\quad \mu\nu} = {1\over 2m} \left [
\partial^\mu F_{\kappa\tau}^{\quad\nu}
- \partial^\nu F_{\kappa\tau}^{\quad \mu} \right ] \, .
\end{eqnarray}  
The constraints are re-written to
\begin{eqnarray}
&&{1\over m} \partial_\mu G_\kappa^{\quad\mu} = 0\, ,\quad
{1\over m} \partial_\mu F_{\kappa\tau}^{\quad\mu} =0\, ,\\
&& {1\over m} \epsilon_{\alpha\beta\nu\mu} \partial^\alpha
T_\kappa^{\quad\beta\nu} = 0\, ,\quad
{1\over m} \epsilon_{\alpha\beta\nu\mu} \partial^\alpha
R_{\kappa\tau}^{\quad\beta\nu} = 0\, .
\end{eqnarray}
 
However, we need to make symmetrization over these two sets
of indices $\{ \alpha\beta \}$ and $\{\gamma\delta \}$. The total
symmetry can be ensured if one contracts the function
$\Psi_{\{\alpha\beta
\} \{\gamma \delta \}}$ with {\it antisymmetric} matrices
$R^{-1}_{\beta\gamma}$, $(R^{-1} \gamma^5 )_{\beta\gamma}$ and
$(R^{-1} \gamma^5 \gamma^\lambda )_{\beta\gamma}$, and equate
all these contractions to zero (similar to the $s=3/2$ case
considered in Ref.~\cite[p. 44]{Lurie}. We obtain
additional constraints on the tensor field functions:
\begin{eqnarray}
&& G_\mu^{\quad\mu}=0\, , \quad G_{[\kappa \, \mu ]}  = 0\, , \quad
G^{\kappa\mu} = {1\over 2} g^{\kappa\mu} G_\nu^{\quad\nu}\, ,
\label{b1}\\
&&F_{\kappa\mu}^{\quad\mu} = F_{\mu\kappa}^{\quad\mu} = 0\, , \quad
\epsilon^{\kappa\tau\mu\nu} F_{\kappa\tau,\mu} = 0\, ,\\
&& T^{\mu}_{\quad\mu\kappa} =
T^{\mu}_{\quad\kappa\mu} = 0\, ,\quad
\epsilon^{\kappa\tau\mu\nu} T_{\kappa,\tau\mu} = 0\, ,\\
&& F^{\kappa\tau,\mu} = T^{\mu,\kappa\tau}\, ,\quad
\epsilon^{\kappa\tau\mu\lambda} (F_{\kappa\tau,\mu} +
T_{\kappa,\tau\mu})=0\, ,\\
&& R_{\kappa\nu}^{\quad \mu\nu}
= R_{\nu\kappa}^{\quad  \mu\nu} = R_{\kappa\nu}^{\quad\nu\mu}
= R_{\nu\kappa}^{\quad\nu\mu}
= R_{\mu\nu}^{\quad  \mu\nu} = 0\, , \\
&& \epsilon^{\mu\nu\alpha\beta} (g_{\beta\kappa} R_{\mu\tau,
\nu\alpha} - g_{\beta\tau} R_{\nu\alpha,\mu\kappa} ) = 0\, \quad
\epsilon^{\kappa\tau\mu\nu} R_{\kappa\tau,\mu\nu} = 0\, .\label{f1}
\end{eqnarray}  
Thus, we  encountered with
the well-known difficulty of the theory of spin-2 particles in
the Minkowski space.
We explicitly showed that all field functions become to be equal to
zero.
Such a situation cannot be considered as a satisfactory one (because it
does not give us any physical information), and it can be corrected in
several
ways.\footnote{The reader can compare our results of this Section with
those of Ref.~\cite{MS}. I became aware about their consideration from
Dr. D. V. Ahluwalia (personal communications, May 5, 1998). I consider their discussion
of the standard formalism in the Sections I and II, as insufficient.}

\section{The Generalized Bargmann-Wigner Formalism for Spin 2.}

We shall modify the formalism in the spirit of  Ref.~\cite{Dvo97}.
The field function (\ref{bwff}) is now presented as
\begin{equation}
\Psi_{\{\alpha\beta\}\gamma\delta} =
\alpha_1 (\gamma_\mu R)_{\alpha\beta} \Psi^\mu_{\gamma\delta} +
\alpha_2 (\sigma_{\mu\nu} R)_{\alpha\beta} \Psi^{\mu\nu}_{\gamma\delta}
+\alpha_3 (\gamma^5 \sigma_{\mu\nu} R)_{\alpha\beta}
\widetilde \Psi^{\mu\nu}_{\gamma\delta}\, ,
\end{equation}
with
\begin{eqnarray}
&&\Psi^\mu_{\{\gamma\delta\}} = \beta_1 (\gamma^\kappa R)_{\gamma\delta}
G_\kappa^{\quad\mu} + \beta_2 (\sigma^{\kappa\tau} R)_{\gamma\delta}
F_{\kappa\tau}^{\quad\mu} +\beta_3 (\gamma^5 \sigma^{\kappa\tau}
R)_{\gamma\delta} \widetilde F_{\kappa\tau}^{\quad\mu} \, ,\\
&&\Psi^{\mu\nu}_{\{\gamma\delta\}} =\beta_4 (\gamma^\kappa
R)_{\gamma\delta} T_\kappa^{\quad\mu\nu} + \beta_5 (\sigma^{\kappa\tau}
R)_{\gamma\delta} R_{\kappa\tau}^{\quad\mu\nu} +\beta_6 (\gamma^5
\sigma^{\kappa\tau} R)_{\gamma\delta}
\widetilde R_{\kappa\tau}^{\quad\mu\nu} \, ,\\
&&\widetilde \Psi^{\mu\nu}_{\{\gamma\delta\}} =\beta_7 (\gamma^\kappa
R)_{\gamma\delta} \widetilde T_\kappa^{\quad\mu\nu} + \beta_8
(\sigma^{\kappa\tau} R)_{\gamma\delta}
\widetilde D_{\kappa\tau}^{\quad\mu\nu}
+\beta_9 (\gamma^5 \sigma^{\kappa\tau} R)_{\gamma\delta}
D_{\kappa\tau}^{\quad\mu\nu} \, .
\end{eqnarray}
 
Hence, the function $\Psi_{\{\alpha\beta\}\{\gamma\delta\}}$
can be expressed as a sum of nine terms:
\begin{eqnarray}
&&\Psi_{\{\alpha\beta\}\{\gamma\delta\}} =
\alpha_1 \beta_1 (\gamma_\mu R)_{\alpha\beta} (\gamma^\kappa
R)_{\gamma\delta} G_\kappa^{\quad\mu} +\alpha_1 \beta_2
(\gamma_\mu R)_{\alpha\beta} (\sigma^{\kappa\tau} R)_{\gamma\delta}
F_{\kappa\tau}^{\quad\mu} + \nonumber\\
&+&\alpha_1 \beta_3 (\gamma_\mu R)_{\alpha\beta}
(\gamma^5 \sigma^{\kappa\tau} R)_{\gamma\delta} \widetilde
F_{\kappa\tau}^{\quad\mu} +
+ \alpha_2 \beta_4 (\sigma_{\mu\nu}
R)_{\alpha\beta} (\gamma^\kappa R)_{\gamma\delta} T_\kappa^{\quad\mu\nu}
+\nonumber\\
&+&\alpha_2 \beta_5 (\sigma_{\mu\nu} R)_{\alpha\beta}
(\sigma^{\kappa\tau}
R)_{\gamma\delta} R_{\kappa\tau}^{\quad \mu\nu}
+ \alpha_2
\beta_6 (\sigma_{\mu\nu} R)_{\alpha\beta} (\gamma^5 \sigma^{\kappa\tau}
R)_{\gamma\delta} \widetilde R_{\kappa\tau}^{\quad\mu\nu} +\nonumber\\
&+&\alpha_3 \beta_7 (\gamma^5 \sigma_{\mu\nu} R)_{\alpha\beta}
(\gamma^\kappa R)_{\gamma\delta} \widetilde
T_\kappa^{\quad\mu\nu}+
\alpha_3 \beta_8 (\gamma^5
\sigma_{\mu\nu} R)_{\alpha\beta} (\sigma^{\kappa\tau} R)_{\gamma\delta}
\widetilde D_{\kappa\tau}^{\quad\mu\nu} +\nonumber\\
&+&\alpha_3 \beta_9
(\gamma^5 \sigma_{\mu\nu} R)_{\alpha\beta} (\gamma^5 \sigma^{\kappa\tau}
R)_{\gamma\delta} D_{\kappa\tau}^{\quad \mu\nu}\, .
\label{ffn1}
\end{eqnarray}
The corresponding dynamical
equations are given by\footnote{All indices in this formula are
already pure vectorial and have nothing to do with
previous notation. The coefficients $\alpha_i$ and $\beta_i$
may, in general, carry some dimension.}
  \begin{eqnarray}
&& {2\alpha_2
\beta_4 \over m} \partial_\nu T_\kappa^{\quad\mu\nu} +{i\alpha_3
\beta_7 \over m} \epsilon^{\mu\nu\alpha\beta} \partial_\nu
\widetilde T_{\kappa,\alpha\beta} = \alpha_1 \beta_1
G_\kappa^{\quad\mu}\,, \label{b}\\
&&{2\alpha_2 \beta_5 \over m} \partial_\nu
R_{\kappa\tau}^{\quad\mu\nu} +{i\alpha_2 \beta_6 \over m}
\epsilon_{\alpha\beta\kappa\tau} \partial_\nu \widetilde R^{\alpha\beta,
\mu\nu} +{i\alpha_3 \beta_8 \over m}
\epsilon^{\mu\nu\alpha\beta}\partial_\nu \widetilde
D_{\kappa\tau,\alpha\beta} - \nonumber\\
&-&{\alpha_3 \beta_9 \over 2}
\epsilon^{\mu\nu\alpha\beta} \epsilon_{\lambda\delta\kappa\tau}
D^{\lambda\delta}_{\quad \alpha\beta} = \alpha_1 \beta_2
F_{\kappa\tau}^{\quad\mu} + {i\alpha_1 \beta_3 \over 2}
\epsilon_{\alpha\beta\kappa\tau} \widetilde F^{\alpha\beta,\mu}\,, \\
&& 2\alpha_2 \beta_4 T_\kappa^{\quad\mu\nu} +i\alpha_3 \beta_7
\epsilon^{\alpha\beta\mu\nu} \widetilde T_{\kappa,\alpha\beta}
=  {\alpha_1 \beta_1 \over m} (\partial^\mu G_\kappa^{\quad \nu}
- \partial^\nu G_\kappa^{\quad\mu})\,, \\
&& 2\alpha_2 \beta_5 R_{\kappa\tau}^{\quad\mu\nu} +i\alpha_3 \beta_8
\epsilon^{\alpha\beta\mu\nu} \widetilde D_{\kappa\tau,\alpha\beta}
+i\alpha_2 \beta_6 \epsilon_{\alpha\beta\kappa\tau} \widetilde
R^{\alpha\beta,\mu\nu}
- {\alpha_3 \beta_9\over 2} \epsilon^{\alpha\beta\mu\nu}
\epsilon_{\lambda\delta\kappa\tau} D^{\lambda\delta}_{\quad \alpha\beta}
= \nonumber\\
&=& {\alpha_1 \beta_2 \over m} (\partial^\mu F_{\kappa\tau}^{\quad \nu}
-\partial^\nu F_{\kappa\tau}^{\quad\mu} ) + {i\alpha_1 \beta_3 \over 2m}
\epsilon_{\alpha\beta\kappa\tau} (\partial^\mu \widetilde
F^{\alpha\beta,\nu} - \partial^\nu \widetilde F^{\alpha\beta,\mu} )\, .
\label{f}
\end{eqnarray} 
The essential constraints are:
\begin{eqnarray}
&&\alpha_1 \beta_1 G^\mu_{\quad\mu} = 0\, ,\quad \alpha_1
\beta_1 G_{[\kappa\mu]} = 0 \,,  \\
&&\nonumber\\
&&2i\alpha_1 \beta_2 F_{\alpha\mu}^{\quad\mu} +
\alpha_1 \beta_3
\epsilon^{\kappa\tau\mu}_{\quad\alpha} \widetilde F_{\kappa\tau,\mu} =
0\,,\\
&&\nonumber\\
&&2i\alpha_1 \beta_3 \widetilde F_{\alpha\mu}^{\quad\mu}
+ \alpha_1 \beta_2
\epsilon^{\kappa\tau\mu}_{\quad\alpha} F_{\kappa\tau,\mu} = 0\,,\\
&&\nonumber\\
&& 2i\alpha_2 \beta_4 T^{\mu}_{\quad\mu\alpha} -
 \alpha_3 \beta_{7}
\epsilon^{\kappa\tau\mu}_{\quad\alpha} \widetilde T_{\kappa,\tau\mu}
= 0\,,\\
&&\nonumber\\
&& 2i\alpha_3 \beta_{7} \widetilde
T^{\mu}_{\quad\mu\alpha} -
\alpha_2 \beta_4 \epsilon^{\kappa\tau\mu}_{\quad\alpha}
T_{\kappa,\tau\mu} = 0\,,\\
&&\nonumber\\
&& i\epsilon^{\mu\nu\kappa\tau} \left [ \alpha_2 \beta_6 \widetilde
R_{\kappa\tau,\mu\nu} + \alpha_3 \beta_{8} \widetilde
D_{\kappa\tau,\mu\nu} \right ] + 2\alpha_2 \beta_5
R^{\mu\nu}_{\quad\mu\nu}  + 2\alpha_3
\beta_{9} D^{\mu\nu}_{\quad \mu\nu}  = 0\,,\\
&&\nonumber\\
&& i\epsilon^{\mu\nu\kappa\tau} \left [ \alpha_2 \beta_5 R_{\kappa\tau,
\mu\nu} + \alpha_3 \beta_{9} D_{\kappa\tau, \mu\nu} \right ]
+ 2\alpha_2 \beta_6 \widetilde R^{\mu\nu}_{\quad\mu\nu}
+ 2\alpha_3 \beta_{8} \widetilde D^{\mu\nu}_{\quad\mu\nu}  =0\,,\\
&&\nonumber\\
&& 2i \alpha_2 \beta_5 R_{\beta\mu}^{\quad\mu\alpha} + 2i\alpha_3
\beta_{9} D_{\beta\mu}^{\quad\mu\alpha} + \alpha_2 \beta_6
\epsilon^{\nu\alpha}_{\quad\lambda\beta} \widetilde
R^{\lambda\mu}_{\quad\mu\nu} +\alpha_3 \beta_{8}
\epsilon^{\nu\alpha}_{\quad\lambda\beta} \widetilde
D^{\lambda\mu}_{\quad \mu\nu} = 0\,,\\
&&\nonumber \\
&&2i\alpha_1 \beta_2 F^{\lambda\mu}_{\quad\mu} - 2 i \alpha_2 \beta_4
T_\mu^{\quad\mu\lambda} + \alpha_1 \beta_3
\epsilon^{\kappa\tau\mu\lambda}
\widetilde F_{\kappa\tau,\mu} +\alpha_3 \beta_7
\epsilon^{\kappa\tau\mu\lambda} \widetilde T_{\kappa,\tau\mu} =0\,,\\
&&\nonumber\\
&&2i\alpha_1 \beta_3 \widetilde F^{\lambda\mu}_{\quad\mu} - 2 i \alpha_3
\beta_7 \widetilde T_\mu^{\quad\mu\lambda} + \alpha_1 \beta_2
\epsilon^{\kappa\tau\mu\lambda} F_{\kappa\tau,\mu} +\alpha_2
\beta_4 \epsilon^{\kappa\tau\mu\lambda}  T_{\kappa,\tau\mu} =0\,,\\
&&\nonumber\\
&&\alpha_1 \beta_1 (2G^\lambda_{\quad\alpha} - g^\lambda_{\quad\alpha}
G^\mu_{\quad\mu} ) - 2\alpha_2 \beta_5 (2R^{\lambda\mu}_{\quad\mu\alpha}
+2R_{\alpha\mu}^{\quad\mu\lambda} + g^\lambda_{\quad\alpha}
R^{\mu\nu}_{\quad\mu\nu}) +\nonumber\\
&+& 2\alpha_3 \beta_9
(2D^{\lambda\mu}_{\quad\mu\alpha} + 2D_{\alpha\mu}^{\quad\mu\lambda}
+g^\lambda_{\quad\alpha} D^{\mu\nu}_{\quad\mu\nu})+
2i\alpha_3 \beta_8 (\epsilon_{\kappa\alpha}^{\quad\mu\nu}
\widetilde D^{\kappa\lambda}_{\quad\mu\nu} -
\epsilon^{\kappa\tau\mu\lambda} \widetilde D_{\kappa\tau,\mu\alpha}) -
\nonumber\\
&-& 2i\alpha_2 \beta_6 (\epsilon_{\kappa\alpha}^{\quad \mu\nu}
\widetilde R^{\kappa\lambda}_{\quad\mu\nu} -
\epsilon^{\kappa\tau\mu\lambda} \widetilde R_{\kappa\tau,\mu\alpha})
= 0\,,\\
&&\nonumber\\
&& 2\alpha_3 \beta_8 (2\widetilde D^{\lambda\mu}_{\quad\mu\alpha} + 2
\widetilde D_{\alpha\mu}^{\quad\mu\lambda} +g^\lambda_{\quad\alpha}
\widetilde D^{\mu\nu}_{\quad\mu\nu}) - 2\alpha_2 \beta_6 (2\widetilde
R^{\lambda\mu}_{\quad\mu\alpha} +2 \widetilde
R_{\alpha\mu}^{\quad\mu\lambda} + \nonumber\\
&+& g^\lambda_{\quad\alpha} \widetilde
R^{\mu\nu}_{\quad\mu\nu}) +
+ 2i\alpha_3 \beta_9 (\epsilon_{\kappa\alpha}^{\quad\mu\nu}
D^{\kappa\lambda}_{\quad\mu\nu}  - \epsilon^{\kappa\tau\mu\lambda}
D_{\kappa\tau,\mu\alpha} ) -\nonumber\\
&-& 2i\alpha_2 \beta_5
(\epsilon_{\kappa\alpha}^{\quad\mu\nu} R^{\kappa\lambda}_{\quad\mu\nu}
- \epsilon^{\kappa\tau\mu\lambda} R_{\kappa\tau,\mu\alpha} ) =0\,,\\
&&\nonumber\\
&&\alpha_1 \beta_2 (F^{\alpha\beta,\lambda} - 2F^{\beta\lambda,\alpha}
+ F^{\beta\mu}_{\quad\mu}\, g^{\lambda\alpha} - F^{\alpha\mu}_{\quad\mu}
\, g^{\lambda\beta} ) - \nonumber\\
&-&\alpha_2 \beta_4 (T^{\lambda,\alpha\beta}
-2T^{\beta,\lambda\alpha} + T_\mu^{\quad\mu\alpha} g^{\lambda\beta} -
T_\mu^{\quad\mu\beta} g^{\lambda\alpha} ) +\nonumber\\
&+&{i\over 2} \alpha_1 \beta_3 (\epsilon^{\kappa\tau\alpha\beta}
\widetilde F_{\kappa\tau}^{\quad\lambda} +
2\epsilon^{\lambda\kappa\alpha\beta} \widetilde F_{\kappa\mu}^{\quad\mu}
+
2 \epsilon^{\mu\kappa\alpha\beta} \widetilde
F^\lambda_{\quad\kappa,\mu})
-\nonumber\\
&-& {i\over 2} \alpha_3 \beta_7 ( \epsilon^{\mu\nu\alpha\beta}
\widetilde
T^{\lambda}_{\quad\mu\nu} +2 \epsilon^{\nu\lambda\alpha\beta} \widetilde
T^\mu_{\quad\mu\nu} +2 \epsilon^{\mu\kappa\alpha\beta} \widetilde
T_{\kappa,\mu}^{\quad\lambda} ) =0\, .
\end{eqnarray}
 
They are  the results of contractions of the field function (\ref{ffn1})
with six antisymmetric matrices, as above. Furthermore,
one should recover the relations (\ref{b1}-\ref{f1}) in the particular
case when $\alpha_3 = \beta_3 =\beta_6 = \beta_9 = 0$ and
$\alpha_1 = \alpha_2 = \beta_1 =\beta_2 =\beta_4
=\beta_5 = \beta_7 =\beta_8 =1$.

As a discussion, we note that in such a framework we have
physical
content because only certain combinations of field functions
can be equal to zero. In general, the fields
$F_{\kappa\tau}^{\quad\mu}$, $\widetilde F_{\kappa\tau}^{\quad\mu}$,
$T_{\kappa}^{\quad\mu\nu}$, $\widetilde T_{\kappa}^{\quad\mu\nu}$, and
$R_{\kappa\tau}^{\quad\mu\nu}$,  $\widetilde
R_{\kappa\tau}^{\quad\mu\nu}$, $D_{\kappa\tau}^{\quad\mu\nu}$,
$\widetilde
D_{\kappa\tau}^{\quad\mu\nu}$ can  correspond to different physical
states
and the equations above describe couplings one state with another.

Furthermore, from the set of equations (\ref{b}-\ref{f}) one
obtains the {\it second}-order equation for the symmetric traceless tensor
of
the second rank ($\alpha_1 \neq 0$, $\beta_1 \neq 0$):
\begin{equation} {1\over m^2} \left [\partial_\nu
\partial^\mu G_\kappa^{\quad \nu} - \partial_\nu \partial^\nu
G_\kappa^{\quad\mu} \right ] =  G_\kappa^{\quad \mu}\, .\label{geq}
\end{equation}
After the contraction in indices $\kappa$ and $\mu$ this equation is
reduced to
\begin{eqnarray}
&&\partial_\mu G_{\quad\alpha}^{\mu} = F_\alpha\, ,  \\
&&{1\over m^2} \partial_\alpha F^\alpha = 0\, ,
\end{eqnarray}
i.~e.,  to the equations connecting the analogue of the energy-momentum
tensor and the analogue of the 4-vector potential (the additional notoph 
field as opposed to the Logunov theory?). 
As we showed in our recent work~\cite{Dvo97} the longitudinal potential 
may have importance
in the construction of electromagnetism (see also the works on the
notoph and notivarg concept~\cite{Tybor}).  Moreover, according to the Weinberg
theorem~\cite{WEIN0} for massless particles it is the gauge part of
the
4-vector potential $\sim \partial_\mu \chi$, which is the physical field. 
The case, when the longitudinal potential is equated to zero, can be considered 
as a particular case only.  This case may be relevant to some physical
situation but hardly to be considered as a basis for unification.
Further investigations may provide additional foundations to
``surprising" similarities of gravitational and electromagnetic
equations in the low-velocity limit, Refs.~\cite{Wein2,Jef,Logunov,Rodrigues}.

\section{Interactions with Fermions.}

The possibility of terms as  $\mathbf\sigma\cdot [ {\mathbf A} \times {\mathbf A}^\ast ]$ appears to be related to the matters
of chiral interactions~\cite{DVO-UNUSUAL,DVO-APS}. As we are now convinced, the Dirac field operator 
can be always
presented as a superposition of the self- and anti-self charge conjugate field operators (cf.~Ref.~\cite{Ziino}). The
anti-self charge conjugate part can give the self charge conjugate part after multiplying by
the $\gamma^5$ matrix, and {\it vice versa}. We derived\footnote{The anti-self charge conjugate field function  $\psi_2$ 
can also be used. The equation has then the form:
\begin{equation}
[ i\gamma^\mu D^\ast_\mu + m ]\psi^a_2 = 0 \, .
\end{equation}
}
\begin{equation}
[ i\gamma^\mu D^\ast_\mu - m ]\psi^s_1 = 0 \, ,\label{31}
\end{equation}
or\footnote{The self charge conjugate field function $\psi_1$ also can be used. 
The equation has the form:
\begin{equation}
[ i\gamma^\mu D_\mu + m ]\psi^s_1 = 0 \, .
\end{equation}
As readily seen, in the cases of alternative choices we have opposite "charges" in the terms of the
type $\mathbf\sigma\cdot [ {\mathbf A} \times {\mathbf A}^\ast ]$ and in the mass terms.}
\begin{equation}
[ i\gamma^\mu D_\mu - m ]\psi^a_2 = 0\,.\label{32}
\end{equation}
Both equations lead to the terms of interaction such as ${\mathbf\sigma}\cdot [ {\mathbf A} \times {\mathbf A}^\ast ]$ provided that the
4-vector potential is considered as a complex function(al). In fact, from (\ref{31}) we have:
\begin{eqnarray}
i\sigma^\mu \nabla_\mu \chi_1 -  m\phi_1 &=& 0\,,\\
i\tilde\sigma^\mu \nabla_\mu^\ast \phi_1 -  m\chi_1 &=& 0\,.
\end{eqnarray}
And, from (\ref{32}) we have
\begin{eqnarray}
i\sigma^\mu \nabla_\mu^\ast \chi_2 -  m\phi_2 &=& 0\,, \\
i\tilde\sigma^\mu \nabla_\mu \phi_2 -  m\chi_2 &=& 0\,.
\end{eqnarray}

The meanings of $\sigma^\mu$ and $\tilde\sigma^\mu$ are obvious from the definition of $\gamma$ matrices. 
The derivatives are defined:
\begin{equation}
D_\mu =\partial_\mu -ie\gamma^5 C_\mu +eB_\mu\,,\quad
\nabla_\mu =\partial_\mu -ieA_\mu\,,
\end{equation}
and $A_\mu =C_\mu +iB_\mu$. Thus, relations with the magnetic monopoles can also be established.

From the above system we extract the terms as $\pm e^2 \sigma^i \sigma^j A_i A^\ast_j$, 
which lead to the discussed terms~\cite{DVO-UNUSUAL,DVO-APS}.\footnote{I am grateful to Prof. S. Esposito 
for the e-mail communications (1997-98) on the alternative proof of the considered interaction.
We would like to note that  the terms of the type ${\mathbf\sigma}\cdot [ {\mathbf A} \times {\mathbf A}^\ast ]$ can be reduced
to $({\mathbf\sigma}\cdot {\mathbf\nabla}) V$, where $V$ is the scalar potential.}
Furthermore, one can come to the same conclusions not applying to the constraints on the
creation/annihilation operators (which we have  previously chosen for clarity and simplicity 
in Ref.~\cite{DVO-APS}). 
It is possible to work with self/anti-self charge conjugate fields and the Majorana
{\it anzatzen}.
Thus, in the considered cases it is the $\gamma^5$ transformation which distinguishes various
field configurations (helicity, self/anti-self charge conjugate properties etc) in the coordinate
representation.

\section{Boson Interactions.}

The most general relativistic-invariant Lagrangian for the symmetric 2nd-rank tensor is
\begin{eqnarray}
\lefteqn{ {\cal L} = - \alpha_1 (\partial^\alpha G_{\alpha\lambda}) (\partial_\beta G^{\beta\lambda})
-\alpha_2 (\partial_\alpha G^{\beta\lambda}) (\partial^\alpha G_{\beta\lambda})}-\nonumber\\
&-&\alpha_3 (\partial^\alpha G^{\beta\lambda}) (\partial_\beta G_{\alpha\lambda})
+m^2 G_{\alpha\beta} G^{\alpha\beta}\,.\label{lagran}
\end{eqnarray} 
It leads to the equation
\begin{equation}
\left [ \alpha_2 \partial^2 +m^2 \right ] G^{\left \{\mu\nu\right\}} + (\alpha_1 +\alpha_3) \partial^{\left \{\mu \vert\right.}
(\partial_\alpha G^{\left.\alpha \vert \nu \right\}})=0\,.
\end{equation}
In the case $\alpha_2 =1  > 0$ and $\alpha_1+\alpha_3=-1$ it coincides with Eq. (\ref{geq}).
There is no any problem to obtain the dynamical invariants for the fields of the spin 2 from the above Lagrangian.
The mass dimension of $G^{\mu\nu}$ is $[energy]^1$.

We now present possible relativistic interactions of the symemtric 2nd-rank tensor. 
They should be the following ones:
\begin{eqnarray}
{\cal L}^{int}_{(1)} &\sim& G_{\mu\nu} F^\mu F^\nu\,,\\
{\cal L}^{int}_{(2)}  &\sim& (\partial^\mu G_{\mu\nu}) F^\nu\,,\\
{\cal L}^{int}_{(3)} &\sim& G_{\mu\nu} (\partial^\mu F^\nu)\,.
\end{eqnarray}
The term $\sim (\partial_\mu G^\alpha_{\,\,\,\alpha}) F^\mu$ vanishes due to the constraint
of tracelessness. Obviously, these interactions cannot be obtained from the free Lagrangian
(\ref{lagran}) just by the covariantization of the derivative $\partial_\mu \rightarrow \partial_\mu + g F_\mu$.

It is also interesting to note that thanks to the possible terms 
\begin{equation}
V (F) =\beta_1 (F_\mu F^\mu) + \beta_2 (F_\mu F^\mu)(F_\nu F^\nu)
\end{equation}
we can give the mass to the $G_{00}$ component of the spin-2 field. This is due to
the possibility of the Higgs spontaneous symmetry breaking~\cite{Higgs}
\begin{equation}
F^\mu (x) =\pmatrix{v+\partial_0 \chi (x)\cr g^1\cr g^2\cr g^3\cr}\,,
\end{equation}
with $v$ being the vacuum expectation value, $v^2 =(F_\mu F^\mu)= -\beta_1/2\beta_2  > 0$.
Other degrees of freedom of the 4-vector field are removed since they can be interpreted as the Goldstone 
bosons. It was stated that ``for any continuous
symmetry which does not preserve the ground state, there is a massless degree of freedom 
which decouples at low energies. This mode is called the Goldstone (or Nambu-Goldstone) particle
for the symmetry". As usual, the Higgs mechanism and the Goldstone modes  should be important in
giving masses to the three vector bosons.\footnote{It is interesting to note the following statement (given without references in wikipedia.org): ``In general, the phonon is effectively the 
Nambu-Goldstone boson for spontaneously broken Galilean/Lorentz symmetry. However, in contrast to the case of internal symmetry breaking, when spacetime symmetries are broken, the order parameter need not be a scalar field, but may be a tensor field, and the corresponding independent massless modes may now be fewer than the number of spontaneously broken generators, because the Goldstone modes may now be linearly dependent among themselves: e.g., the Goldstone modes for some generators might be expressed as gradients of Goldstone modes for other broken generators."}
As one can easily see, this expression does not permit an arbitrary phase for $F^\mu$, which is  possible only if 
the 4-vector would be the complex one.

Next, due to the Lagrangian interaction of fermions with notoph  are of the order $e^2$ since the beginning
(as opposed to the interaction with the 4-vector potential $A_\mu$), it is more difficult to observe it.
However, as far as I know the theoretical precision calculus in QED (the Land\'e factor, the anomalous magnetic moment, the hyperfine splittings in positronium and muonium, and the decay rate of o-Ps and p-Ps)  are about the order corresponding to the 4th-5th loops,
where the difference may appear with the experiments~\cite{DFT,Kinoshita}.

\section{Conclusions.}

We considered the Bargmann-Wigner formalism to derive the equations for the AST field and for 
the symmetric tensor of the 2nd rank.
We introduced additional scalar normalization field in the Bargmann-Wigner formalism in order 
to take into account possible physical
significance of the Ogievetski\u{\i}-Polubarinov--Kalb-Ramond modes. We introduced the additional symmetric matrix in the
Bargmann-Wigner expansion $(\gamma^5 \sigma^{\mu\nu} R)$ in order to take into account the dual fields.
The consideration is similar to Ref.~\cite{DVO-FS}.

Furthermore, we discussed the interactions of notoph, photon and graviton (and, probably, 
notivarg\footnote{In order to analize its dynamical invariants and interactions
one should construct the Lagrangian from the analogs of the Riemann tensor $\widetilde D^{\mu\nu,\alpha\beta}$.}).
For instance, the interaction notoph-graviton may give the mass to spin-2 particles in the way which is similar to 
the spontaneous-symmetry-breaking Higgs formalism.

\section*{Acknowledgements}
I am grateful to the referee of ``International Journal of Modern Physics" and
``Foundation of Physics", whose advice of the mass dimensions
(normalizations) of the fields  was very useful.
I acknowledge insightful discussions with participants of recent conferences
on Symmetries and the Standard Model.  I am thankful to organizers and {\it El Colegio Nacional}
for partial financial support during the QTS-8. I am thankful to the UAEM for 
partial financial support  during the INRACOM13.

%% The bibliography section

\end{document}